# Supporting Intel® SGX on Multi-Package Platforms[1]

Simon Johnson, Raghunandan Makaram, Amy Santoni, Vinnie Scarlata

Intel Corporation

{simon.p.johnson, makaram.raghunandan, amy.santoni, vincent.r.scarlata}@intel.com

**ABSTRACT**

Intel® Software Guard Extensions (SGX) was originally released on client platforms and later extended to single socket server platforms. As developers have become familiar with the capabilities of the technology, the applicability of this capability in the cloud has been tested. Various Cloud Service Providers (CSPs) are demonstrating the value of using SGX based Trusted Execution Environments (TEE) to create a new paradigm of Confidential Cloud Computing.

This paper describes the additional platform enhancements we believe are necessary to deliver a user programmable Trusted Execution Environment that scales to cloud usages, performs and is secure on multi-package platforms.

## 1 Introduction

In introducing Intel® SGX [1], Intel has changed the way applications can provide better protection to its own secrets without having to fully rely on the underlying and more privileged platform software. Usages that benefit from this type of protection are wide ranging, including protection of fingerprint matching, password manager vault protection, blockchain wallets and smart-contract verification, digital rights management for premium content protection, and a variety of other key protection usages.

In recent years as applications become more dependent on services from the cloud and the network edge, more and more data is being processed on platforms that do not necessarily belong to the data owner or the service provider. This places cloud and network edge operators in a unique place, where they want to host other parties' workloads, but do not want access to those workloads or their data. Most of these hardware owners would like to tell their customers – we'll host your workloads, but you don't have to trust us to not reveal your data either intentionally or unintentionally through security breaches.

The birth of the confidential computing paradigm, using Trusted Execution Environments to provide protection to data and the CSP's reputation is unsurprising.

To deliver this paradigm a Trusted Execution Environment that scales and performs is required. Using desktop PC's and mobile chips only deliver the possibility in concept. What we require is a TEE that can scale to many concurrent instances, running on many cores on platforms with more than one CPU socket.

### 1.1 Scaling SGX Requirements and Challenges

To address this need, Intel looked to expand SGX from client to server. The main outcome we wanted to achieve was that the software development experience and API remained unchanged so that SGX enabled application software, which is platform agnostic, should just work whether it's running on Intel® Core i7 client or an Intel® Xeon Series server.

To achieve this outcome, we had to address challenges that were specific to SGX.

1. Keys used for attestation and sealing are derived from per CPU package secrets. We need to make the platform, consisting of multiple CPU packages cryptographically coherent with a single identity.
2. Memory residing in protected ranges, physically attached to one package, must be securely accessible from another package.

At the same time we were looking at these challenges the advent of containerizing whole applications inside an SGX enclave [2] [3] started to arise as a serious usage scenario. This challenged the view that applications would only use small amounts of protected memory, consequently this led us to address two other significant challenges:

1. Reduce the cost of additional performance implications that the Memory Encryption Engine [4] introduces to workload owners.
2. Reduce the memory storage overhead costs that platform owners need to pay for the memory integrity and replay protection tree.

We looked at several ways to help reduce both the size of the integrity tree and the impact on performance created by the additional memory access of the integrity tree. Balancing these questions results in a performance vs security trade-off discussion.

---

## 2 Creating a Single Identity

### 2.1 Overview

On single package platforms, such as standard clients, SGX functionality and security properties are provided completely by the one package. Each package ships with per-part unique keys built into the HW. SGX instructions allow enclaves to access keys derived from these HW keys to help protect secrets or securely communicate between enclaves. Unique signing keys can also be derived. Using these signing keys, along with certificates issued by Intel, 3rd parties can remotely authenticate that they are communicating with a genuine SGX processor package. The signing keys can be used to provision attestation keys to the platform or be used directly for attestation. These keys are common across all cores on a single package. A key request will result in the same answer on all the cores in the package. Operating system schedulers rely on this coherency. Losing this property would require complex operating system changes.

Establishing a single coherent software environment on multi-socket platforms creates several new platform requirements.

1. Keys available to the enclave must be consistent when a process is scheduled on different packages. Additionally, this requirement should not be met by allowing a package's hardware keys (effectively its identity) be exposed outside that package. Under this restriction a new set of platform-wide shared keys must be established.
2. Keys generated in the field and used as a foundation for attestation must be registered with the attestation infrastructure to be recognized during attestation key provisioning.
3. When a package is moved in to a different platform, it must not continue to use the source platform's keys.
4. Exposing existing platform keys to new package requires that package first to be vetted by the infrastructure that certified the platform with an attestation key.

### 2.2 Multi-Package Life-cycle Stages.

Deploying a multi-package platform has 2 stages: Platform Establishment, and Platform Registration.

Platform Establishment generates new platform-wide keys along with an authenticated manifest that describes all the devices with access to those keys. Next, Platform Registration makes use of the manifest to register platform and its components with a Registration Service. The Registration Service authenticates the manifest and the devices referenced in it. The Registration Service generates the certificates necessary for attestation key provisioning. After registration, these certificates allow the existing provisioning infrastructure to recognize the new platform the same as it does a standard client.

#### 2.2.1 Platform Establishment

During Platform Establishment, packages configure the platform with common keys and secure inter-package communication.

The processor package contains unique, per-part hardware keys, called the SGX root keys. User data is then encrypted using derivatives of those keys.

Figure 1 shows the SGX data protection model when 2 packages are used. Rather than encrypting user data with the hardware keys in the packages, derivatives of new "platform" root keys are used. Data can then be accessed regardless of which physical package. To ensure the platform keys are accessible after the platform is reset, they are encrypted and stored in persistent storage (ex. flash). Each package encrypts its own copy of the platform keys using its hardware keys. This ensures that if any package fails, the remaining packages can still access the platform keys.

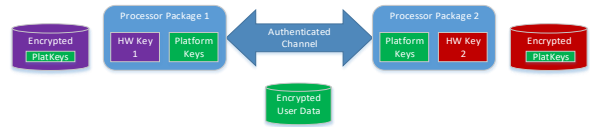

**Figure 1: Multi-package SGX Keys**

Packages use unprotected memory as a communication medium to negotiate secure session keys. Using these sessions, the packages compare/agree on configurations and generate the new platform keys. If a physical link connects two packages, they will also negotiate and configure a set of link protection keys.

#### 2.2.2 Platform Registration

The Registration Service bridges the attestation provisioning gap between client provisioning model and multi-package platform provisioning.

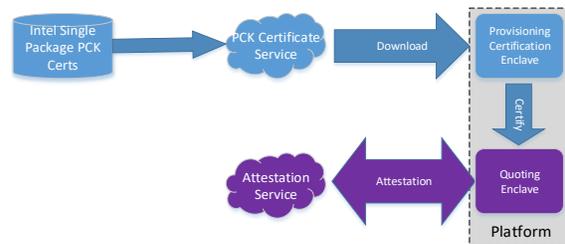

**Figure 2: Single Package Provisioning**

Figure 2 depicts the single package (E3) server attestation model based on Intel® Data Center Attestation Primitives (DCAP) services [5]. Each package has a unique TCB-based identity key called the Provisioning Certification Key (PCK). Intel published certificates for all PCKs in a PCK Certificate Service. During server or VM initialization, this certificate is downloaded by platform software. In most cloud deployments, the CSP will host a local PCK certificate caching service that will download certificates from the Intel® PCK Certificate Service in bulk and then make them available to CSP servers without any runtime communication with Intel. The Provisioning Certification Enclave uses the PCK to certify Attestation Keys created by Quoting Enclaves on the platform.

Attestation Services, customers, or other challengers of the platform can request Quotes from the platform, which now have a certificate chain back to Intel.

The PCK is derived from the SGX Provisioning Root key, which is in fuses. Each processor has a Provisioning Root Key and a Sealing Root Key. Intel stores a copy of the Provisioning Root Key and uses its knowledge of the fuse keys to derive the package's PCK and issues a PCK certificate for the processor.

On a multi-package platform, during Platform Establishment, platform root keys are generated randomly. Unlike clients, this may take place in the field. A registration step is required for Registration Service to be able to derive the keys this platform will use and in turn issue certificates for them.

The Registration Service evaluates the packages and configuration of the platform using the manifest. If the Registration Service approves of the platform as a trustworthy SGX environment, it uses copies of the Platform Provisioning Root Keys creates enclosed in the manifest to create certificates for the PCKs derived from the platform's Provisioning Root Key. Once created, the PCK certificates are delivered to the PCK Certificate Service.

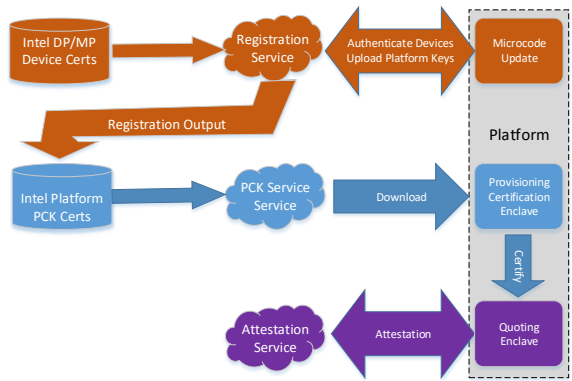

**Figure 3: Multi-Package Provisioning**

Since the Platform Provisioning Root Keys are exposed to the Registration Server, an SGX platform can only have one Registration Server for a platform at a time. Otherwise we would need to attest to both Registration Servers. The only way to change Registration Servers is to delete the platform keys and create a new instance that can be registered with the different Registration Service.

## 3 Performant Memory Protections

### 3.1 Memory Protection Performance

In order to expand the amount of memory covered by any one Memory Encryption Engine (MEE) [4], there are two methods which can be used to increase the size of memory protected.

1. Increase the amount of on-die storage used for the top-level anti-replay counters
2. Add more levels to the anti-replay tree

Expanding the amount of on-die storage only increases the amount of protected memory linearly. Adding a level on the other hand increases the protected memory by a factor of 8 (using the current design). To support large memory EPC sizes, this consumption of memory is onerous.

Numerous studies have shown that workloads that make significant use of SGX memory have their performance impacted. The reason for this is the additional bandwidth required to fetch the various levels of the integrity tree protecting the SGX memory.

While caching and other techniques can be used to address some of this performance loss we concluded that it was best to take a different approach to Multi-Package memory in order to significantly boost the performance of SGX on these types of platform. Removing the additional memory access for the integrity tree reduces the performance impact to a small latency on each memory operation due to encryption.

The next few sections discuss the security benefits of using an integrity replay tree and what additional changes we required in order to ensure critical benefits are still maintained with the new approach.

### 3.2 Physical Memory Protections are Orthogonal from Access Controls

Implementation of SGX only relies on the use of an inline memory encryption engine to provide protection to code/data as it leaves the CPU package. It does not use encryption to provide separation between enclaves and regular memory (when resident in caches for instance) nor does it rely on encryption to provide separation between enclaves - they are all encrypted using the same key. The SGX architecture implements several access controls which prevents access to cache lines that are resident inside the package in decrypted form. Changing the physical memory protection scheme does not change these access controls.

We should also note that SGX was not designed with a single form of memory protection in mind. Indeed, if you look at how the CPU enumerates the protected memory regions through CPUID.SGX_LEAF (EAX=12h), then you can see that there is sufficient flexibility built-in.

#### 3.2.1 What software attacks does the MEE provide protection against?

While the MEE provides enhanced protections against hardware-based attacks on memory, its architecture also provides additional protections in the SW domain, Table 1 shows these.

Since we were also looking at providing a regular form of memory encryption under the control of the Virtual Machine Manager (VMM), expanding the Multi-Key Total Memory Encryption (MK-TME) [Ref TME] architecture to add a separate key for SGX on the platform would resolve a number of these issues particularly the Reset and EPC Reclaim based attacks. The largest remaining concern would be memory aliasing issues as these would allow simple SW based replay and integrity attacks .

| Attack | Description | Resulting loss |
|---|---|---|
| Reset | The attacker forces a reboot of the platform and forces the previous memory range protections to be dropped. Secrets can now be retrieved from uninitialized memory. | Confidentiality |
| Aliasing/ EPC Replay | Attacker configures a second system address to map to a protected memory location. Ability to replay memory of arbitrary data allows code injection into enclave. | Execution Control / Confidentiality |
| EPC Reclaim | If platform has ability to tear down protections post use (without reset) EPC protected secrets would be exposed. | Confidentiality |
| DIMM Config. Attacks | SW attacks DIMM configuration settings to prevent writes from becoming persistent. | Integrity / Confidentiality |

**Table 1: MEE Attack Protection**

## 3.3 Memory Aliasing

Aliasing occurs when two addresses can map to a single physical data location. In some cases, like page tables, this is entirely allowable, and the Intel Architecture relies on this trick to allow remapping of operating system pages in application address spaces. However, when dealing with security, an alias could be used to bypass the SGX memory range checks that occur as part of the access control mechanisms used for enforcing isolation of enclave memory from other software.

The memory encryption engine with its integrity/anti-replay scheme caught changes to SGX memory pages that may have occurred through an alias when the line was correctly consumed. The MK-TME engine does not have these protections so we must prevent SW aliases in the SGX memory region; otherwise malicious SW can cause bad things to occur.

Given the complexity of server platforms, there are several sources of aliasing. These include memory configuration architecture itself and Reliability Availability and Serviceability (RAS) feature known as DIMM sparing.

There are two types of alias that could occur:

1. Aliasing within EPC region (inside-in)
2. Aliasing from outside into the EPC region (outside-in)

### 3.3.1 Inside-In Aliasing
In the case of inside-in, two system addresses (SA3 & SA4 in Figure 4) in the SGX protected region of memory alias to the same physical address. This allows a malicious enclave running in the protected region to bypass enclave ownership checks to gain access to any other enclave resident in the EPC page affected.

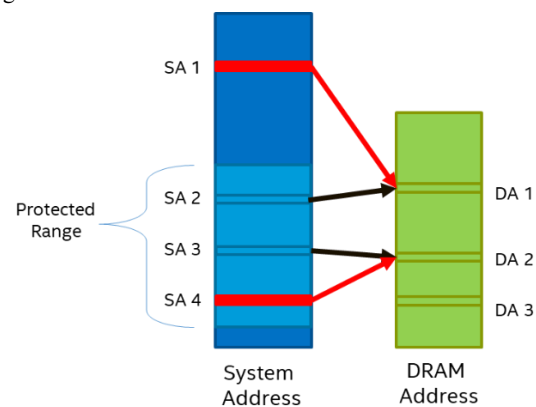

**Figure 4: Aliasing**

### 3.3.2 Outside-In Aliasing
In the case of outside-in aliasing a system address (SA1) outside the protected region memory aliases with another system address (SA2) to the same region of protected memory, effectively bypassing the SGX EPC range check protections. This case can be used by untrusted software outside the enclave to manipulate any enclave resident in the EPC page affected.

Eliminating Aliases are therefore critical to ensuring that software integrity and anti-replay protections.

## 3.4 Preventing Aliasing

### 3.4.1 Outside PRM → Inside PRM
Memory used in servers uses a special form of memory that has extra storage for metadata which are used for features which improve the Reliability, Availability and Serviceability of the platform. This is typically known as Error Correct Code memory (ECC-memory). To resolve the outside-in aliasing issue, we purpose one of the bits in the DRAM ECC metadata to indicate whether the line previously written was an SGX specific line.

In the cases where a regular memory access is used to read a line that was previously written via secure path, a fixed value is returned. This would prevent attacks which rely on knowing just cipher text values. In the case where a secure memory read encounters a line was previously written in insecure mode, the system determines that the enclave is under attack. This will result in the disabling of SGX. Note that the HW checks for outside-in aliases is performed after RAS ECC detection and correction. This ensures that DRAM soft and hard errors corrected by ECC will not lead to disabling of SGX.

| Read Secure | Written Secure | Returns |
|---|---|---|
| No | No | Cache line |
| No | Yes | Fixed Value |
| Yes | No | Fixed Value & Poison |
| Yes | Yes | Cache line |

**Table 2: Detecting Outside-In Attacks**

#### 3.4.2  Inside PRM → Inside PRM

Unfortunately, there is no hardware we can add to support the run-time detection of inside-in aliases. Instead, during system boot, all the locations inside the SGX EPC need to be checked to ensure they do not overwrite each other. Conveniently a piece of trusted FW runs during the SGX initialization process which undertakes certain memory configuration and lock checks before SGX can be enabled. This FW is used to perform checks to ensure there are no aliases within the PRMRR region.

### 3.5  Comparison of Memory Protections

In summary, you can now see with the changes that we have in place for datacenter security, SW adversaries remain outside the trust boundary for SGX whilst balancing performance required for many scalable server workloads.

| Protect Memory from | Attack Vector | SGX | SGX-TEM |
|---|---|---|---|
| Loss of Confidentiality | SW | Yes | Yes |
| Loss of Integrity | SW | Yes | Yes |
| Anti-Replay | SW | Yes | Yes |
| Loss of Confidentiality | HW | Yes | Yes* |
| Loss of Integrity | HW | Yes | No |
| Anti-Replay | HW | Yes | No |

**Table 3: Memory Protections Comparison**

*The final thing to note is that SGX-TEM relies on MKTME for confidentiality. The cryptographic scheme used can only mitigate a class of HW attacks where the adversary can only see the cipher text once and not while the system is changing the data.

## 4  Coherent Memory Protections

Intel® Xeon® Multi CPU package servers use a high-speed interconnect for keeping memory coherent between the packages. While this link is proprietary, it is exposed externally on the package and is therefore open to attack.

### 4.1  Memory Coherency Architecture Overview

In Figure 5 you can see a basic block diagram of the major components involved in memory coherency in a two-socket system with a single Intel® Ultra Path Interconnect (UPI) Link between them.

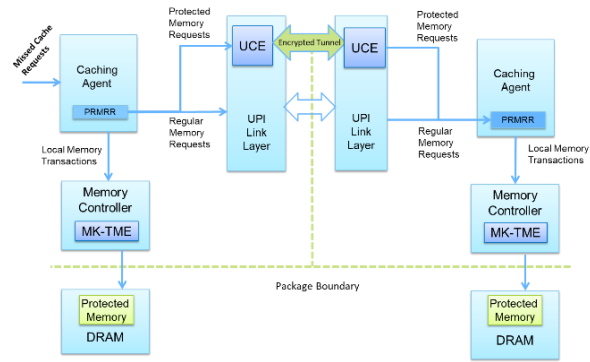

**Figure 5: Memory Coherence Architecture**

When a core on the first package is referencing a memory location and it fails to locate that line in its local cache, the reference is passed on to the local Caching Agent. These agents understand which physical memory locations are attached to which CPU package (via BIOS configuration process) and forwards the request to the relevant UPI interconnect. Memory locations that are considered secure are sent with a secure attribute set. The UPI link agent forwards requests with the secure attribute to UPI Crypto Engine (UCE) for protection prior to transmission. Non-secure requests are forwarded in plain-text to the appropriate queue for transmission. UCE will provide confidentiality, integrity and replay protection for the secure requests and data using a counter-mode based scheme before transmission.

Upon receipt in the receiving package, the UPI agent determines whether the packets making up the message need decrypting. After decrypting the UCE will forward the message to its local Caching Agent with the secure attribute set. The local caching agent checks that memory requests for SGX memory have the secure attribute set, before issuing memory references are issued to the local memory controller for the secure memory.

Memory location cache lines are returned to the remote caching agent before being forwarded to the UPI for secure transmission back to the request originating socket.

## 4.2 Attacks on SGX Memory Coherency

Table 4 identifies the attacks that can occur on the memory coherency architecture.

| Threat | Description |
|---|---|
| Eavesdropping | Attacker passively sniffs packets for SGX memory resulting in confidentiality loss |
| Tampering | Attacker modifies SGX packets to break the coherency model for the SGX memory |
| Replay | Attacker forges/replays a SGX packet to break the coherency model for the SGX memory |
| Directory Corruption | Attacker corrupts the directory bits in DRAM corresponding to SGX memory |
| Misconfiguration | Attacker manipulates configuration assets that affect the security of the SGX memory. |

**Table 4: SGX Memory Coherency Threats**

To mitigate a number of these threats (i.e. Eavesdropping, Tampering and Replay), selecting the correct cryptographic protections is very important. In this case, a counter-mode based encryption scheme with integrity should suffice.

To prevent misconfiguration attacks each package must check these configurations for consistency and lock the configuration before SGX can be enabled.

## 5 Platform Configuration

The architectural enhancements described so far support the creation of a single key and the enabling of memory communications encryption. This section covers the configuration aspects of the architecture which need to occur during platform set-up.

The process overall is fairly simple. BIOS determines the enablement scenario the platform is in during boot and prepares the platform accordingly. Then, it triggers an Intel FW module that runs on each CPU package. The FW module executes the FW module that performs the following sequence:

1. Checks Platform configuration consistency and locked state across packages
2. Calculates/decrypts and programs keys in HW used for identity, data protection, bus protection, and memory protection
3. Performs Memory Alias checking for Inside-in memory scenario

## 5.1 Supported Configuration Scenarios

The Intel FW supports the following main scenarios:

1. Establish New Platform
2. Reboot Old Platform
3. Add Package to Platform

### 5.1.1 Establishing a New Platform

During Platform Establishment, a new platform instance is created along with a corresponding set of platform keys and material necessary to register the newly created instance.

Packages start by establishing authenticated communication channels over unprotected memory. These channels are used to share the new platform keys. The packages create a signed manifest of the pairings and platform keys.

Each package creates an encrypted key structure containing a copy of the platform keys, the communication keys from each pairing, and platform configurations that are not permitted to change.

For each UPI link, packages program link encryption keys and the platform is ready to boot with SGX.

New key structures and manifests are returned to BIOS for future use.

### 5.1.2 Rebooting a Platform

If the BIOS determines that every package has a key structure with keys for the currently loaded microcode update, the BIOS instructs Microcode Update to conduct a standard SGX boot.

During a standard boot, previously created keys are decrypted and loaded; sessions between packages are re-established; UPI buses are configured with encryption keys; and SGX is ready to be activated.

### 5.1.3 Adding a New Package to a Platform

Once the platform instance has been established and secrets are provisioned, unverified devices must not have access to that instance's keys.

Attestations on the platform are rooted in the Registration Service-issued PCK certificate. This is an assertion that the platform composition was evaluated and approved. The same Registration Service must ensure that that assertion remains true if the platform is using those keys.

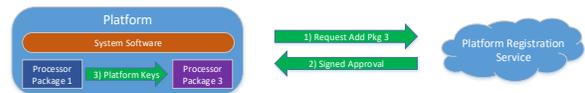

**Figure 6: Flow for Adding New Package**

When a new package is found, BIOS creates an Add Request identifying the platform and new package. System SW sends the request to the Registration Service. If the request is approved, the service creates a Platform Membership Certificate, which other packages in the system will use to authenticate the new package. After the next reset, existing package(s) verify the certificate and share the previously established platform keys with the new package.

## 5.2 Key Exchange

Platform establishment requires secure communication between packages before inter-package HW encryption over UPI is enabled.

To establish these connections, packages use unprotected memory as a channel to conduct signed Diffie-Hellman based key negotiations using their PRKs. The resulting keys are the long-lived Master Comms Keys. All packages negotiate keys with their adjacent packages and a BIOS appointed coordinator package called the Master Package. Some steps are taken by the master package on behalf of the entire platform.

On every reset, Master Comms Keys are used to establish encryption/MAC keys between packages. Using these keys, processors exchange and verify consistency of memory configurations and information found in the Platform Info structure.

If inconsistencies are found SGX is not enabled.

## 6 Summary

To extend the SGX architecture to support multi CPU-package systems such as Intel® Xeon Scalable Processor based platforms, we identified a method for creating a single identity for an SGX platform from CPU HW that is manufactured separately through the creation of a registration process. In addition, we demonstrated how SGX related memory coherency traffic is protected as it transported between CPU packages on high speed package interconnect. Finally, we discussed memory protection from SW can be maintained and performance increased by adding additional checks in the memory architecture using ECC metadata.

Further information about SGX can be found at http://software.intel.com/sgx .